\documentclass[aps,prl,twocolumn,showpacs,amsmath,amssymb,floatfix,superscriptaddress]{revtex4-1}

\usepackage{graphicx}
\usepackage[bookmarks,bookmarksopen,bookmarksnumbered,colorlinks,linkcolor=red,linktocpage,citecolor=blue,urlcolor=cyan,pdfpagemode=UseOutline]{hyperref}

\def\n{n}
\def\sss{\scriptscriptstyle\rm}
\def\ben{\begin{equation}}
\def\een{\end{equation}}
\def\xc{_{\sss XC}}
\def\br{{\bf r}}
\def\half{\frac{1}{2}}
\def\Hx{_{\sss HX}}

\newcommand{\parref}[1]{(\ref{#1})}
\newcommand{\intd}{\mathrm{d}}
\providecommand{\abs}[1]{\left|#1\right|}
\DeclareMathOperator{\tr}{tr}

\def\wt{\mathtt{w}}

\begin{document}

\title{Direct extraction of excitation energies from ensemble density-functional theory}
\author{Zeng-hui Yang}
\affiliation{Microsystem and Terahertz Research Center, China Academy of Engineering Physics, Chengdu, China 610200}
\author{Aurora Pribram-Jones}
\affiliation{Department of Chemistry, University of California, Berkeley, CA 94720, USA}
\affiliation{Lawrence Livermore National Laboratory, Livermore CA 94550, USA}
\author{Kieron Burke}
\affiliation{Department of Chemistry, University of California-Irvine, Irvine, CA 92697, USA}
\author{Carsten A. Ullrich}
\affiliation{Department of Physics and Astronomy, University of Missouri, Columbia, MO 65211, USA}
\date{\today}

\begin{abstract}
A very specific ensemble of ground and excited states is shown to yield an exact formula for
any excitation energy as a simple correction to the energy difference between orbitals of
the Kohn-Sham ground state.
This alternative scheme avoids either the
need to calculate many unoccupied levels as in time-dependent density
functional theory (TDDFT) or the need for many self-consistent ensemble
calculations.  The symmetry-eigenstate Hartree-exchange (SEHX) approximation
yields results comparable to standard TDDFT for atoms.
With this formalism, SEHX yields approximate
double-excitations, which are missed by adiabatic TDDFT.
\end{abstract}

\maketitle

The Hohenberg-Kohn (HK) theorem \cite{HK64,L79,L82,L83} of ground-state
density-functional theory (DFT) \cite{HK64,KS65} has several parts.
The most-used in practice is the establishment of an exact density
functional, $F[\n]$, whose minimum yields the exact ground-state
density and energy of a given system.  Almost all practical calculations
use the Kohn-Sham (KS) scheme \cite{KS65} to minimize $F$ with an approximation to the small
exchange-correlation contribution, $E\xc[\n]$.  In fact, many properties
of interest in a modern chemical or materials calculation can be extracted
from knowledge of the ground-state energy as a function of nuclear
coordinates, or in response to a perturbing field.

However, except under very special circumstances, most optical excitation
frequencies cannot be deduced.  Hence there has always been interest in
extending ground-state DFT to include such excitations.  Moreover,
another part of the HK theorem guarantees that such frequencies (and all
properties) are indeed functionals of the ground-state density.
In recent years, linear-response
time-dependent DFT (TDDFT) \cite{RG84,C96,MMNG12,U12,UY14} has become a popular route for extracting
low-lying excitation energies of molecules, because of its unprecedented
balance of accuracy with computational speed \cite{M16}.  For significantly sized
molecules, more CPU time will be expended on a geometry optimization than
a single TDDFT calculation on the optimized geometry.

However, while formally exact,
TDDFT with standard approximations is far from perfect. If the unknown
exchange-correlation (XC) kernel of TDDFT is approximated by its zero-frequency
(and hence ground-state) limit, no multiple excitations survive \cite{M16}.
While a useful work-around exists for cases where a double is close to a
single excitation \cite{MZCB04,CZMB04},
there is as yet no simple and efficient general procedure for
extracting double excitations within adiabatic TDDFT \cite{Elliott2012}.

Ensemble DFT (EDFT) \cite{T79,GOKb88} applies the principles of ground-state DFT to a convex
ensemble of the lowest $M$ levels of a system, for which a KS system can be defined \cite{GOK88}.
EDFT is formally exact, but practical
calculations require approximations, and initial attempts yielded
disappointing
results \cite{OGKb88}.  Accuracy is greatly
improved when so-called ``ghost interactions'' between distinct states are removed from the approximations \cite{GPG02}.  EDFT remains
an active research area because, being variational, it should not
suffer from some of the limitations of standard TDDFT.  Recent strides
by Pernal and coworkers \cite{PGP13,Pastorczak2014},
Fromager and coworkers \cite{SKJF15,Deur2017}, and others attempt to create a useful practical
alternative to TDDFT, but the difficulty remains in finding accurate
low-cost approximations. EDFT usually requires running several different self-consistent
ensemble calculations to extract
several low-lying excitations.

Here we (a) derive a formula from EDFT to correct a KS orbital
energy difference into an exact excitation energy, without doing any
self-consistent ensemble calculations, (b) argue that its computational
cost should typically be less than either standard TDDFT or EDFT, (c) calculate this
correction using the symmetry-eigenstate Hartree-exchange (SEHX) approximation \cite{PYTB14,YTPB14,supplemental} for atoms, demonstrating its
accuracy relative to standard TDDFT, and (d) show that SEHX estimates double excitations.

EDFT is a formally exact and variational
excited-state method \cite{T79,GOKb88,GOK88}.   Let $E_i$ be the electronic energy levels, $i=0,1,...$,
each with degeneracy $g_i$.  Construct an ensemble from positive convex weights $w_i$,
letting $I$ be the maximum non-zero weight. The weights are not variational parameters. Then, from the foundational theorems,
the ensemble energy
\ben
E_{I}(\{\wt_i\})=\sum_{i=0}^{I} g_i\, \wt_i\, E_i,  ~~~\sum_{i=0}^{I} g_i\wt_i=1,
\een
is a functional of the ensemble density
\ben
n^\text{ens}(\br)=\sum_{i=0}^{I}\wt_i\, \tilde{n}_i(\br),
\een
where $\tilde{n}_i(\br)$ is the sum of all densities in the $i$th multiplet (so that $\int d^3r\, \tilde{n}_i(\br)= g_i N$, with N being the number of electrons),
and can be found via a minimization, so long as the weights are
monotonically non-increasing.   Applying the same conditions to
a fictitious system of non-interacting electrons with the same
weights, one can define a KS system whose ensemble density
matches the interacting one.  Defining energy components in
the usual way, only the XC contribution needs to be approximated
to perform an ensemble DFT calculation.  Since the ensemble energies
depend linearly on the weights (at least, in the exact theory), one
can easily deduce transition frequencies.

\def\GOK{^{\rm GOK}}
Infinitely many ensembles can be realized, but the GOK ensemble from the
original work \cite{GOK88} is particularly useful and popular, in which
all weights are the same except for the highest multiplet, i.e.,
\ben
\wt_{i\neq I}= (1-g_I \wt)/M_{I-1},~~~~~
\wt_I=\wt ,~~~~({\rm GOK})
\label{eqn:GOKweights}
\een
where $M_I$ is
the number of states up to and including the $I$-multiplet,
and $\wt \leq M_I^{-1}$ to preserve convexity.
When $\wt=M_I^{-1}$, the weights are all equal (an equiensemble).
In general, the corresponding ensemble
density must be found by self-consistent solution of the ensemble KS
equations, for the given weights.  The excitation energy of the
$I$ multiplet can only be isolated by performing self-consistent
calculations for all lower multiplets.
The excitation energy from the ground state
to the $I$th multiplet is \cite{GOK88}
\ben
\omega_I=\frac{1}{g_I}\left.\frac{d E\GOK_{I}}{d \wt}\right|_{\wt_I}+
\sum_{i=1}^{I-1}\frac{1}{M_i}\left.\frac{d E\GOK_{i}}{d \wt}\right|_{\wt_i},
\label{eqn:GOKexciteng}
\een
requiring $I+1$ self-consistent calculations, including
the ground-state KS calculation, where the density is held fixed when
the derivative is taken.

The weights defined by Eq. \parref{eqn:GOKweights} are also a
linear interpolation between two consecutive equiensembles, containing
$M_{I-1}$ and $M_I$ states.  Thus
$\omega_I$ can also be calculated via
\ben
\omega_I=M_I\, E_{I}(\wt=M_I^{-1})-M_{I-1}\, E_{I-1}(\wt=M_{I-1}^{-1}),
\label{eqn:subtractensemble}
\een
which requires only two self-consistent calculations.
However, if one needs all excitation energies up to $\omega_I$,
$I+1$ self-consistent calculations are still needed.
The computational costs of Eqs. \parref{eqn:GOKexciteng} and
\parref{eqn:subtractensemble} are much higher
than TDDFT with the Casida equation \cite{C96}.

\def\GOKII{^{\rm GOKII}}
Now we reintroduce an alternative one-parameter ensemble, in which all
states have weight $\wt$ {\em except the ground state}:
\ben
\wt_0=\frac{1-\wt(M_I-g_0)}{g_0},~~\wt_{i\neq 0}=\wt.~~~~{\rm (GOKII)}
\label{eqn:footnoteweights}
\een
We say reintroduce, as this ensemble was mentioned in a footnote in
Ref. \cite{GOK88}, although never applied (as far as we know).
However, we can show (see supplemental material \cite{supplemental})
that the excitation energy using Eq. \parref{eqn:footnoteweights} has
a much simpler formula than using Eq. \parref{eqn:GOKexciteng}:
\ben
\omega_I=
\frac{1}{g_I}\left[\left.\frac{d E\GOKII_{I}}{d \wt}\right|_{\wt_I}
-\left.\frac{d E\GOKII_{I-1}}{d \wt}\right|_{\wt_{I-1}}\right].
\label{eqn:footnoteexciteng}
\een
Despite the simplicity, in general
one still needs to do $I+1$ calculations to get all excitation energies.
However, unlike Eq. \parref{eqn:GOKweights}, the set
of weights defined by Eq. \parref{eqn:footnoteweights} is now a
linear interpolation between the {\em ground state} and the equiensemble of
$M_I$ states. Now, $\wt=0$ recovers the
ground state, not an equiensemble with one less multiplet.
A further simplification is made by
noting that the EDFT formalism is valid even as $\wt\to 0$.
Setting $\wt_I=\wt_{I-1}=0$ in
Eq. \parref{eqn:footnoteexciteng} and defining
$\Delta\omega_I=\omega_I-\omega_I^\text{KS}$, where
$\omega_I^\text{KS}$ is the KS orbital energy difference,
yields
\ben
\Delta\omega_I=\frac{1}{g_I}
\frac{d}{d\wt}\Bigg|_{\wt=0}\left( E\GOKII_{\sss{XC},I}
- E\GOKII_{\sss{XC},I-1}\right),~~~~({\rm DEC})
\label{eqn:w0exciteng}
\een
where
$E\GOKII_{\sss{XC},I}$ is the GOKII
ensemble XC energy functional \cite{GOK88}
containing up to the $I$th multiplet.
This is a direct ensemble correction (DEC) to the KS transition
frequencies.

Equation \parref{eqn:w0exciteng} is the central formal result of this work.
Because all elements of the right-hand side are evaluated on the
ground-state density, this correction is a formally exact ground-state
density functional for correcting KS transitions into physical transitions.
If approximated by an explicit density functional, it could be
evaluated at no noticeable additional cost to a standard ground-state
DFT calculation.
Compared with the cubic scaling of the TDDFT linear response equations \cite{C96},
Eq. \parref{eqn:w0exciteng} is vastly more efficient.
On the other hand, TDDFT yields both transition frequencies
and oscillator strengths, as well as dipole overlap matrices.  In addition, linear
response TDDFT can yield spatially resolved response functions,
once perturbations different from a long-wavelength electric field are allowed.
In future work, we will explore what else, beyond transition frequencies,
might be extracted in a manner similar to Eq. \parref{eqn:w0exciteng}.

There is an infinite number of excited-state ensembles.
Even if we consider only those that interpolate between
the ground state and the equiensemble, Eq. \parref{eqn:footnoteweights}
is not the only choice.  The exact ensemble functional yields the same
result in {\em any} ensemble, but approximations yield different results
for different ensembles.  A DEC expression is a particularly simple route
to excitation energies.

Eqs. \parref{eqn:GOKweights} and \parref{eqn:footnoteweights}
are identical for a simple bi-ensemble, the ensemble of
a non-degenerate ground and first excited states.
Studies of $\wt=0$ bi-ensembles have been carried
out previously \cite{L95}, as well as
calculations of the first excitation energy \cite{PYTB14,YTPB14}.
Thus the DEC of Eq. \parref{eqn:w0exciteng} can be viewed as a generalization of
such results to an arbitrary excitation.

The exact $E\xc\GOKII$ of Eq. \parref{eqn:w0exciteng} can
be obtained numerically for simple cases \cite{PYTB14,YTPB14},
but in practical calculations $E\xc\GOKII$ must be approximated.
In general, the $E\xc\GOKII$ of Eq. \parref{eqn:w0exciteng}
must account for the state ordering and differences in multiplet
structure between the real and KS systems, which poses a challenge for the
development of approximations.

SEHX \cite{PYTB14,YTPB14} is an explicit orbital-dependent
ensemble-density functional generalization of
the exact-exchange approximation (EXX) of ground-state DFT, whose full
expression is given in the supplemental material\cite{supplemental}.
Using the energy decomposition of Nagy \cite{N01,N02},
SEHX constructs the combined Hartree-exchange energy from
an ensemble sum over spin- and spatially-symmetrized
multi-determinant KS wavefunctions, removing ``ghost interactions" and
approximating the ensemble discontinuity \cite{L95}, and yielding
good results in the GOK ensemble \cite{PYTB14,YTPB14}.
Inserting SEHX into Eq. \parref{eqn:w0exciteng},
all the contributions from excitations below $I$ cancel, yielding
an approximation that depends only on the difference between a contribution
from the $I$th multiplet and the ground state:
\ben
\Delta\omega_I^\text{SEHX}= H_I/g_I -H_0/g_0.
\label{eqn:w0SEHX}
\een
Here $H_i=H_i^{\rm orb}+H_i^{\rm dens}$, where
\ben
H_i^{\rm orb} =
\half\int\frac{\intd^3r\intd^3r'}{\abs{\br-\br'}}
\tr\left\{{\bf V(\br,\br')}\cdot{\bf Q}_i\right\},
\een
and
\ben
H_i^{\rm dens}=-\int d^3r\, v\Hx(\br)\, \tilde{n}_i(\br).
\een
${\bf V}$ is a matrix containing products of KS orbitals,
${\bf Q}_i$ is a matrix containing orbital occupation factors and
symmetrization coefficients of KS determinants
(see supplemental material \cite{supplemental}), and
$v\Hx(\br)$ is just the ground-state Hartree-exchange potential.
As our tests are on atoms and ions, we use
the KLI approximation \cite{KLI92} for $v\Hx(\br)$ to obtain more accurate
orbital energies than those from semilocal approximations \cite{PGB00}.
We denote calculations with Eq. \parref{eqn:w0SEHX} as DEC/SEHX.
Unlike Eq. \parref{eqn:w0exciteng}, Eq. \parref{eqn:w0SEHX} depends only
on the ground and excited states in question, so the state ordering
problem is bypassed and calculation
is highly efficient.
The ordering-independency of Eq. \parref{eqn:w0SEHX} is due to SEHX, yielding
reasonable excitation energies
even if the KS state ordering is different from the real one.
On the other hand, the approximate state ordering might not be correct.

To illustrate the performance of DEC/SEHX, we calculated excitation
energies of small atoms.  To see exclusively the effect of the
excitation method \cite{KSB13}, we use the exact KS potential and energies
for the He and Be atoms \cite{UG93,UG94}. We compare with TDDFT
using the adiabatic local density approximation (ALDA) \cite{U12}.
For simplicity, we use the Tamm-Dancoff approximation (TDA) \cite{HH99}
in TDDFT calculations, and we checked to make sure that the
results only change slightly with and without TDA.
The results are shown
in Figs. \ref{fig:HeSEHXExact} and \ref{fig:BeSEHXExact}.
More results for atoms are available in the supplemental
material \cite{supplemental}.

\begin{figure}
\includegraphics[width=\columnwidth]{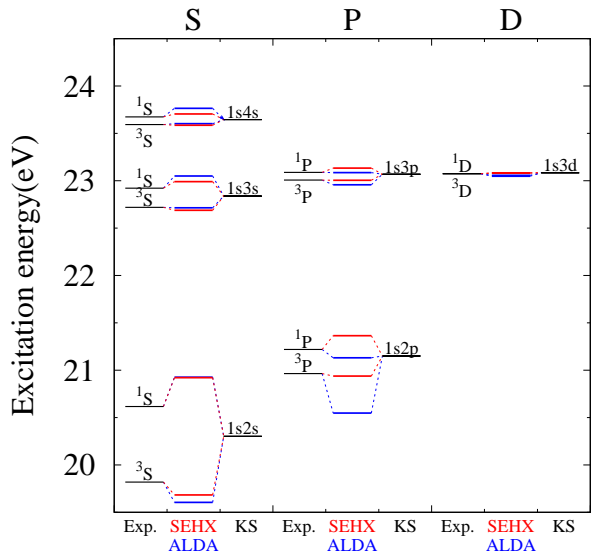}
\caption{Exact KS and true excitations of the He atom (black).  Experimental
values from the NIST atomic spectra database \cite{NIST,MWD06}.
DEC/SEHX excitation energies in red and
TDDFT/ALDA results within TDA in blue.}
\label{fig:HeSEHXExact}
\end{figure}

Figure \ref{fig:HeSEHXExact}  shows the He results.
These are all single excitations (as all doubles
in He are in the continuum).  The DEC/SEHX gives results
that are qualitatively similar to those of standard TDDFT.  In fact, the mean
absolute errors are typically about 30\% {\em smaller}, despite the lack of
approximate correlation in the DEC calculation.

\begin{figure}
\includegraphics[width=\columnwidth]{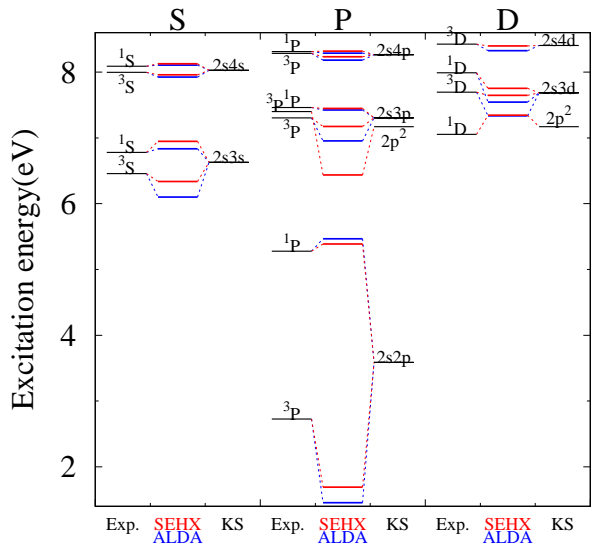}
\caption{Same as Fig. \ref{fig:HeSEHXExact}, but for Be\cite{NIST,KM97}. Configurations denoted
without core.
The $2\mathrm{p}^2$ configuration corresponds to two
doubly-excited states ($3^3\mathrm{P}$ and $1^1\mathrm{D}$).}
\label{fig:BeSEHXExact}
\end{figure}

Figure \ref{fig:BeSEHXExact}  shows the results for Be, again with the
exact KS potential.  For single excitations, the results are quantitatively
similar to those of He, again with DEC errors being noticeably smaller than
their TDDFT/ALDA counterparts.  But in DEC we can also calculate the double
excitations, which are completely absent from any adiabatic TDDFT calculation.
We note that the double excitations are less accurate than their single
counterparts, but since there are only two, this might be incidental.
The supplemental material \cite{supplemental} gives many more atomic calculations, using
approximate ground-state KS potentials, showing the strong sensitivity of
both DEC and TDDFT to the KS levels in atoms.

To better understand the performance of DEC/SEHX for the double excitations,
we turn to a much simpler model problem that was designed to study precisely
this question.   Consider two fermions in a 1d harmonic potential with a
contact interaction \cite{MZCB04,ZB04}:
\ben
\hat{H}=\frac{1}{2}\sum_{i=1}^2 \left(-\frac{\partial^2}{\partial x_i^2}+
x_i^2\right)+\lambda \delta(x-x'),
\label{eqn:Hooke}
\een
where $\lambda > 0$.  For small values of $\lambda$, the system is weakly
interacting, and exchange-type approximations are accurate.

\begin{table}[htbp]
\begin{tabular}{|c|c|c|c|cc|c|}
\hline
\multicolumn{2}{|c|}{} &\multicolumn{5}{c|}{$\Delta\omega_I$}\\
\hline
 & & &TDDFT&\multicolumn{2}{c|}{DEC}&TDDFT\\
$I$ &$\omega_{KS}$ & exact &AEXX & SEHX & exact & dressed\\
\hline
\multicolumn{7}{|c|}{Singles}\\
\hline
1 & 962 & 38 & 39 & 39 & 38 & 39\\
3 & 1953 & 47 & 30 & 30 & 48 & 49\\
5 & 2948 & 52 & 25 & 27 & 51 & 54\\
\hline
\multicolumn{7}{|c|}{Doubles}\\
\hline
2 & 1923 & 41 & -- & 58 & 41 & 39\\
4 & 2915 & 49 & -- & 77 & 49 & 47\\
\hline
\end{tabular}
\caption{Exact and approximate singlet $\Delta\omega_I$ (in mH) of the
1D two-electron contact-interaction Hooke's atom with $\lambda=0.2$.
The dressed TDDFT results are calculated with an exchange-only frequency-dependent kernel \cite{MZCB04}.}
\label{table:Hooke}
\end{table}
The results are shown in Table \ref{table:Hooke}.  Because this is 1d, there
are no degeneracies or multiplets.  However, this model was purposely
constructed to have near-degeneracies between the multiple and single
excitations.  With the harmonic confining potential, as $\lambda\to 0$,
many levels approach one another.  As shown, the double excitation of level
2 is very close to the single of level 3, and the double at level 4 is very close to the single at level 5.

In the 4th and 5th columns of the table, we report exact exchange results.
The former is TDDFT, using the exact KS potential and the exact ground-state
exchange in an adiabatic approximation.  The latter is DEC/SEHX.  We see
that both are excellent approximations to the lowest excitations, and give almost identical
results for the single excitations.  This is because $\lambda=0.2$, ensuring
that correlation effects are relatively weak.  But, unlike adiabatic TDDFT,
DEC/SEHX also yields predictions for the double excitations.  Just like
in the atoms, the errors are substantially larger for the doubles.

Because this model has only two electrons, we can calculate the exact DEC numerically with Eq. \parref{eqn:w0exciteng},
by calculating the exact energies, densities and the xc potential of the model first. We then evaluate Eq. \parref{eqn:w0exciteng} numerically using these exact quantities (see supplemental material \cite{supplemental}). These DEC/exact results are in column 6, and agree within a mH with the
exact results.  This shows that exact DEC does handle doubles correctly,
so that the failing in DEC/SEHX is due to the lack of correlation.  The last
column of the table shows results with the DSPA, a frequency-dependent
model XC kernel designed for weakly-correlated systems with strong coupling
between a single and double excitation, often called
dressed TDDFT \cite{MZCB04,CZMB04}.  This works extremely well here, as this
system was designed to illustrate its accuracy.  Study of the difference
in the results between these two should provide a route to improving
DEC approximations for double excitations.

A discerning reader might have noted that, throughout this work, we have avoided
discussion of $N$- and $v$-representability\cite{PY89,G75,L82}.  These issues have been partially
explored \cite{OGKb88,HT85} within EDFT in general, but not for this particular
ensemble.  But none of the calculations here ran into any representation
difficulties, such as an inability to find a KS system with the required
density.  There is little reason to fear such problems in practice.
Furthermore, as we use only DEC, any such difficulties in EDFT in general
are likely to be least problematic for our applications.

There is obviously much work to be done to see if DEC can become competitive
with standard TDDFT calculations.  It should be applied to molecules with
standard ground-state functionals, to see if the results are as
accurate or if semilocal ground-state approximations
destroy the accuracy found here within SEHX.  Other challenges for TDDFT,
such as charge transfer excitations, should be carefully tested.  In such
a case, we are less hopeful that DEC will provide accurate results as,
like TDDFT, it also begins from (unrelaxed) KS transitions of the
ground-state.  Other ensembles might also yield direct ensemble
corrections, or properties other than simple excitation energies might
be accessible.

Several other EDFT-based methods for excitations
were recently proposed, such as the linear interpolation method \cite{SKJF15},
the Helmholtz free-energy minimization method \cite{PGP13},
and the ensemble-referenced Kohn-Sham method (REKS) \cite{F08}.
The REKS method is a multi-reference extension to ground-state DFT and EDFT (see also \cite{Pastorczak2016}),
while the others are within standard EDFT.
Each has its own advantages, and the REKS method has been
shown to work well in strongly-correlated systems \cite{FHB15}.
However, all these methods require extra self-consistent calculations
aside from the ground-state one. The simplification achieved in this paper
by changing the ensemble type suggests that
similar simplifications may also be possible in these methods. Another route for future research would bypass the use of ensemble functionals altogether by developing approximate methods based on the DEC.

In summary, DEC (in
Eq. \parref{eqn:w0exciteng}), is a formally
exact approach
to excitation energies from DFT, as
illustrated by our model harmonic trap calculation.
For example, where the fundamental and optical gaps match (insulating
solids without excitons), DEC yields a new approach to the problem
of finding accurate gaps within DFT\cite{PYBY17}, relating
the derivative discontinuities with respect to particle number\cite{PPLB82} and those with
respect to optical excitation\cite{L95}.
While DEC and TDDFT are both post-processing steps after a ground-state KS
calculation,
DEC is less expensive and applicable to
traditionally difficult problems such as multiple excitations and
spin-multiplets. Unlike TDDFT, EDFT is based on a variational principle \cite{GOKb88}, so
the DEC derived in this work may be more reliable than TDDFT corrections,
which are based on response theory.
The calculations shown in this paper merely
demonstrate the DEC method: SEHX yields
better accuracy than TDDFT/ALDA for single excitations in atoms, and
approximates doubles (albeit less accurately than singles).
Simpler approximations, avoiding
solution of OEP-type equations, might produce
usefully accurate results for valence excitations in molecules.
Thus DEC represents an exciting alternative to TDDFT.

The authors thank Cyrus Umrigar for providing us the exact KS potentials of the He and Be atoms. Z.-H.Y. is currently supported by Science Challenge Project No. TZ2016003 (China). Z.-H.Y. and C.A.U. were supported by NSF grant DMR-1408904. A.P.J. was supported by the University of California President's Postdoctoral Fellowship. Part of this work was performed under the auspices of the U.S. Department of Energy by Lawrence Livermore National Laboratory under Contract DE-AC52-07NA27344. K.B. was supported by DOE grant DE-FG02-08ER46496.

\end{document}